\documentclass[11pt,preprint2]{aastex}

\newcommand{\simle}{\mbox{$\stackrel{<}{_{\sim}}$}}

\newcommand{\micronn}{\,\hbox{$\mu$m}}

\newcommand\lkha{LkH$\alpha$~101}

\slugcomment{Draft - do not circulate}

\shorttitle{Imaging of \lkha}
\shortauthors{Tuthill et al.}

\begin{document}

\title{Imaging the disk around the luminous young star 
        \lkha\ with infrared interferometry}

\author{ P. G. Tuthill\altaffilmark{1}, J. D. Monnier\altaffilmark{2}, 
W. C. Danchi\altaffilmark{3}, D. D. S. Hale \altaffilmark{4} and 
C. H. Townes\altaffilmark{4}}

\altaffiltext{1}{School of Physics, University of Sydney, NSW 2006, Australia}
\altaffiltext{2}{Smithsonian Astrophysical Observatory MS\#42,
60 Garden Street, Cambridge, MA, 02138}
\altaffiltext{3}{NASA Goddard Space Flight Center,
Infrared Astrophysics, Code 685, Greenbelt, MD 20771}
\altaffiltext{4}{Space Sciences Laboratory, University of California, Berkeley,
Berkeley,  CA  94720-7450}

\email{gekko@physics.usyd.edu.au, jmonnier@cfa.harvard.edu, 
       wcd@snoopy.gsfc.nasa.gov, david@isi.mtwilson.edu, cht@ssl.berkeley.edu}

\begin{abstract}
The Herbig Ae/Be star \lkha\ has been imaged at high angular resolution at
a number of wavelengths in the near-infrared (from 1 $\sim$ 3\micronn) using 
the Keck~1 Telescope, and also observed in the mid-infrared (11.15\micronn)
using the U.C.~Berkeley Infrared Spatial Interferometer (ISI).
The resolved circular disk with a central hole or cavity reported in
\citet{lk_nature} is confirmed.
This is consistent with an almost face-on view (inclination $\simle 35 ^\circ$)
onto a luminous pre- or early-main sequence object surrounded by a massive 
circumstellar disk.
With a multiple-epoch study spanning almost four years, relative motion of 
the binary companion has been detected, together with evidence for changes 
in the brightness distribution of the central disk/star.
The resolution of the \lkha\  disk by ISI mid-infrared interferometry 
constitutes the first such measurement of a young stellar object in this
wavelength region. 
The angular size was found to increase only slowly from
1.6 to 11.15\micronn, inconsistent with standard power-law temperature
profiles usually encountered in the literature, supporting instead
models with a hot inner cavity and relatively rapid transition to
a cool or tenuous outer disk.
The radius of the dust-free inner cavity is consistent with a model of
sublimation of dust in equilibrium with the stellar radiation field. 
Measurements from interferometry have been combined with published photometry 
enabling an investigation of the energetics and fundamental properties
of this prototypical system.
\end{abstract}


\section{Introduction}

Among the brightest young stellar objects in the infrared sky, \lkha\
is thought to be a transitional object which is on (or nearly on) the 
main sequence, but is still surrounded by a massive circumstellar disk.
Following active accretion, hot stars are expected to pass through a 
brief phase in which the remnant accretion disk, which retains up to 
$\sim$0.3 times the mass of the central star \citep{Shu_90,Hollenbach_94}, 
is sculpted and eventually dissipated by the radiation and wind from the 
newborn star.
The structure of these disks has been speculative, with uncertainty
surrounding characteristic sizes and the existence of an inner cavity.

Images of young circumstellar disks in Orion\citep{Odell_93,
McG_O_96} and remnant or fossil disks at large distances from
the star around ``Vega-type'' sources such as $\beta$~Pic \citep{Smith_84} 
have been seen in reflected light, or silhouetted against bright nebulosity.
This work has attracted intense interest for, as first hypothesized 
by \citet{kant}, it is likely that our own solar system grew out of 
such a flattened primordial nebula \citep{Beckwith_96} labelled by
Kant an ``Urnebel''.
 
The framework for understanding stellar formation is based on a 
rapidly-evolving theoretical and observational picture.
Although the formation of circumstellar disks has long been favored, for
the case of massive stars there have been historical problems with getting
models which fit the shape of the spectral energy distribution (SED) 
\citep{Hillenbrand_92} to account for the luminosities in the near-IR
\citep{Hartmann93}, and with forbidden line profiles not matching expectations
\citep{Bohm_Catala_94}.
It was also found that the SED could be fitted by spherically symmetric shells
\citep{Miroshnichenko_97,Pezzuto_97} or composite shell-disk models
\citep{Miroshnichenko_99}.

Recent observational evidence has argued both for and against the presence
of disks. 
Millimeter-wave interferometry has found evidence for large-scale rotation
in Herbig~Ae stars \citep{Mannings_Sargent97,Mannings_Sargent00}, albeit at 
much larger spatial scales than the expected sizes of the inner disks. 
However, similar studies of Herbig~Be stars \citep{Fuente_01} conclude that 
the disks have dissipated at an early stage before the star becomes visible.
Even restricting attention to just Herbig~Ae stars, HST coronographic
observations of AB~Aur \citep{Grady_99} show a more circularly symmetric 
structure consistent with disk viewed pole-on, contrasting with the mm 
observations.

Recent high-resolution imaging in the infrared has also been equivocal.
Classical accretion disks have generally performed poorly in attempting to 
fit observations from the latest generation of long-baseline interferometers
\citep{IOTA_ABAur_99,Raphael_01,Akeson_00}.
More directly, the expected source asymmetries from a population of disks,
most of which must be inclined to the line of sight, were not indicated
\citep{Raphael_01}.
However, of the three systems for which the hot inner regions have been
well resolved by full interferometric imaging in the infrared, 
HK~Tauri~B \citep{koresko98}, \lkha\ \citep{lk_nature} and 
MWC~349 \citep{mwc349}, clear disk morphologies 
(edge-on, face-on and edge-on respectively) have been established.
In contrast to the near-IR where disks have proven larger than expected, 
\citet{Hinz_01} failed to detect extended flux around a sample of Herbig Ae 
stars from mid-IR nulling interferometry.

Modellers are now re-thinking the simple thin/flared-disk geometries with
power-law thermal profiles, now seen by many as inadequate in light of the 
measurements of Millan-Gabet and others, and promising new candidates for 
reconciling some of the discrepancies are emerging.
One new model \citep{Dullemond_01} involves an inner circumstellar cavity 
whose radius is set by thermal dust evaporation which 
adjoins a flaring disk geometry \citep{Chiang_Goldreich_97} at larger radii.
Such flared disks  may have thermal instabilities causing runaway local 
heating, puffed walls and self-shadowing of more distant regions 
\citep{Dullemond_00}.
The resultant geometry is capable of intercepting more of the stellar 
radiation, thus boosting its near-IR excess while maintaining excellent 
fits to the SED.
This has been verified experimentally with new spectral data on four
young stars \citep{Natt_01}.

\lkha was first identified as the source of illumination of the irregular
reflection nebula NGC~1579 by Herbig (1956), who matched the H$\alpha$ 
spectrum with a faint, deeply embedded star lying $\sim$5 arcseconds within
the border of a dark lane crossing the few-arcminute sized nebula.
In the years since this discovery, it has become one of the most studied
young stellar objects with observations spanning the spectrum.
Despite this, pinning down many of its basic properties has proved elusive. 
The 800\,pc distance established by \citet{Herbig_71} was based on 
measurement of two nearby stars believed to be in association.
Recently, \citet{Stine_Oneal_98} made a strong case based on radio 
photometry that the system is much closer; suggesting instead that
it is located in an extension of the Taurus-Auriga star formation 
complex at 160\,pc. 

Part of the difficulty has been the complexity of the highly anisotropic
circumstellar environment, with at least four molecular clouds 
\citep{Redman_86,Barsony_90} surrounding \lkha\ and its associated
H~{\small II} region, S222 \citep{Herbig_56}.
A deeply-embedded population of up to hundreds of probable protostars, 
pre-main-sequence low-mass stars, and brown dwarf candidates has
been detected in the radio \citep{Becker_White_88,Stine_Oneal_98} and
infrared \citep{Barsony_91,Aspin_Barsony_94}.
Estimates of the visible extinction vary wildly in the literature by almost 10 
magnitudes between values $A_v = $ 9.4 \citep{Barsony_90} to 
18.5\,mag \citep{Hou_97}.

This paper presents results from extremely high angular resolution imaging
obtained with interferometric techniques.
This has allowed observation of the hot, young, self-luminous disk surrounding
\lkha\ in the near- and mid-infrared.
The observational methods and the two different instruments used to obtain 
the data are described in Section~\ref{obs}, with the results given in
Section~\ref{results}.
The astrophysical interpretations are discussed in Section~\ref{discussion},
while section Section~\ref{conclusions} contains a summary of the important
findings.

\section{Observations}
\label{obs}

Observations were made at a number of epochs over the interval 1997--2001 
and utilized two separate instruments to secure high resolution imaging
data in the near- and mid-infrared.
Near-IR observations were taken with the Keck-I telescope, while the 
mid-IR interferometry was performed using the U.C.~Berkeley Infrared Spatial 
Interferometer (ISI).
These are briefly described in turn below.

\subsection{Near-Infrared Interferometry} 

Observations at the Keck~I telescope employed the technique of 
aperture masking interferometry in order to recover information out 
to the diffraction limit \citep{keckmask} of the 10\,m primary mirror.
Placing a mask over the telescope pupil has been shown under some
circumstances to confer signal-to-noise advantages over full-pupil
speckle interferometry \citep{HB92}. 
A large number of rapid-exposure, high magnification data frames were
processed to extract Fourier amplitudes and closure 
phases, enabling images to be produced from a self-calibration
algorithm based on the maximum entropy method \citep{Siv_84,SB84}.
A detailed description of the apparatus and data processing can
be found in \citet{keckmask}, and its application in the study of young
stars demonstrated \citep{Bittar01,lk_nature}

An observing log of near-IR observations is given in Table~\ref{tab:nirc} 
showing the dates, filters, and point-source calibrator stars used for each 
data set.
The observing wavelength was selected from the standard complement of 
interference filters in the facility Near-IR Camera (NIRC), which offered 
a range of bandwidths (from about 1 to 20\%) over the near-infrared region
\citep{nirc_96}.
The characteristics of filters used are given in Table~\ref{tab:filters}.
All observations utilized the annular ring shaped pupil \citep{keckmask},
with the exception of 1999~February~05 observation within the J-band, which utilized 
the full unobstructed pupil due to the relatively low apparent luminosity of 
\lkha\ in this spectral region.
An additional difficulty with the J- and some H-band observations was associated 
with the fact that the finest fringes were beyond the Nyquist sampling limit of
the camera.
This prevented recovery of data at a few of the very longest H-band baselines,
while for J-band even intermediate baselines had serious problems with fringe
power aliasing, reducing the effective resolution to that of a 4\,m telescope.

\begin{table}[h]
\centering
\caption{Journal of near-IR Keck observations }
\begin{tabular}{cll}
  &  &  \\
\hline
 Date (UT) & Filter & Calibrator Star \\
\hline
1997 Dec 17 & H     & HD 27349 \\
            & KCONT & HD 27349 \\
            & PAHCS & HD 27349 \\
1997 Dec 19 & H     & HD 27349 \\
            & KCONT & HD 27349 \\
            & PAHCS & HD 27349 \\
1998 Sep 30 & H     & 54 Per   \\
            & BR$\gamma$ & HD 27349 \\
            & CH4   & HD 27349 \\
            & PAHCS & HD 27349 \\
1999 Jan 06 & H     & 54 Per   \\
            & CH4   & 54 Per   \\
1999 Feb 05 & J     & HD 28423 \\
2000 Jan 26 & H     & 54 Per   \\
            & CH4   & 54 Per   \\
2001 Jul 30 & H     & HD 27349 \\
            & CH4   & HD 27349 \\
2001 Sep 04 & J     & HD 28423 \\
            & H     & 54 Per   \\
            & CH4   & HD 27349 \\
            & K     & HD 27349 \\
\hline
\end{tabular}
\label{tab:nirc}
\end{table}

\begin{table}[h]
\centering
\caption{Interference filters used in the near-IR }
\begin{tabular}{lcc}
\hline
Name & $\lambda_0$ & $\delta\lambda$   \\
 & ($\mu$m) & ($\mu$m)  \\
\hline
J      & 1.251 & 0.292 \\
H      & 1.658 & 0.333 \\
BR$\gamma$ & 2.165 & 0.022 \\
K      & 2.214 & 0.427 \\
KCONT  & 2.260 & 0.053 \\
CH4    & 2.269 & 0.155 \\
PAHCS  & 3.083 & 0.101 \\
\hline
\end{tabular}
\label{tab:filters}
\end{table}

\subsection{Mid-Infrared Interferometry} 

Mid-IR visibility data were obtained at 11.15\,$\micron$ with the
ISI, a two-element heterodyne stellar interferometer located
on Mt.~Wilson, CA.
Both telescopes are mounted within movable semi-trailers which allowed 
reconfiguration of the baseline over the course of these measurements.
Periodic observations of K giant stars $\alpha$~Tau and $\alpha$~Boo 
were utilized to calibrate fringe visibilities and to monitor drifts 
to within a few percent accuracy.
Detailed descriptions of the apparatus, observing, and data
reduction procedures can be found in Hale et al. (2000).
A journal of \lkha\ observations taken with the ISI is provided in 
Table~\ref{tab:isidata}. 
Spatial frequencies and position angles listed are averages of
a small range spanned during the course of observations.

\begin{deluxetable}{cccccc}
\tablewidth{0pt}
\tablecaption{Journal of Mid-IR ISI Observations}
\tablehead{
\colhead{Date} & \colhead{Tel Separation} & \colhead{Spatial Frequency} & 
\colhead{Position Angle\tablenotemark{a}} &  \colhead{Visibility} \\ 
 & \colhead{(meters)} & \colhead{($10^{5}$ $\rm{rad}^{-1}$)} &
 \colhead{(Degrees)} & \\
}
\startdata
 1998 Oct &  4.0 &  3.28  & 73  & 0.964 $\pm$ 0.087 \\
 1999 Sep & 13.3 &  11.47 & 106 & 0.648 $\pm$ 0.068 \\
\enddata
\tablenotetext{a}{Position angle is measured in degrees East of North}
\label{tab:isidata}
\end{deluxetable}

\section{Results}
\label{results}

\subsection{Multi-Wavelength Near-Infrared Images}

\begin{figure*}
\begin{center}
\includegraphics[angle=0,width=14cm]{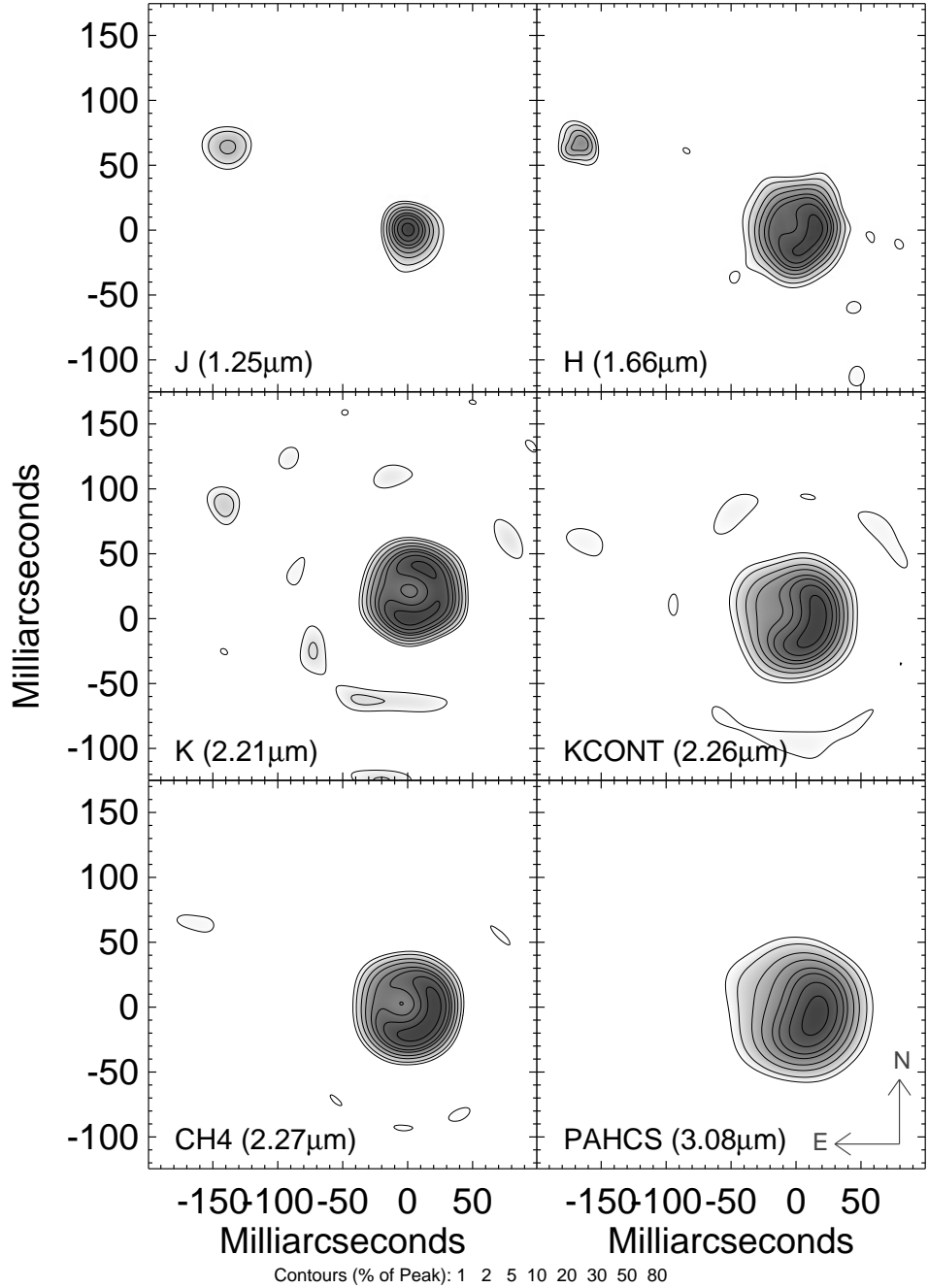} 
\caption{Images of \lkha\ reconstructed from aperture masking 
interferometry data taken over 6 different filter bandpasses.
To enhance the signal-to-noise, most frames depicted are the average
of a number of individual reconstructions taken at various epochs, as
described in the text. 
\label{fig:sixcolour}}
\end{center}
\end{figure*}

Multi-wavelength images of \lkha\ are given in Figure~\ref{fig:sixcolour}.
For the J- and K-band images, a single map produced from data taken in
2001~Sep is shown.
The remaining images are averages of four separate observations over the 
interval 1997~December to 1999~January, with a stronger weighting given to high 
signal-to-noise data obtained when observing conditions were favorable.
Although this averaging of maps could help reduce random noise, care
had to be taken not to include data from widely separate epochs as
secular changes in appearance have been detected in this source, as 
described in more detail below.

The main features presented by this system are a bright resolved 
circular disk and a separate unresolved companion about 180\,mas
to the E-NE.
The resolved component exhibits a central hole or depression in
flux, and a marked asymmetric brightening towards the W-SW limb.
These basic features were reliably reproduced over multiple
epochs under a range of different experimental conditions.

It is important, however, to point out inherent imperfections in this
imaging process.
Spurious structures are present in the images, for example 
the faint circular ring appearing at the lowest contour 
level in the K, KCONT and CH4 maps at a radius of around 90\,mas.
This feature would be familiar to radio astronomers as a `grating ring' 
and is caused by incomplete Fourier sampling.
It should also be emphasized that the images are reconstructions based on
the maximum-entropy technique of a target which is not highly resolved.
There is a very large parameter space of images which would fit to the 
data; the maximum-entropy solution attempts to choose (in some sense) the 
smoothest, most featureless map from this set (see e.g. 
Skilling \& Bryan 1984, Narayan \& Nityananda 1986).
This implicit assumption is not always the best one to make in the context
of astronomical imaging, however for the present we adopt it as the standard 
starting point.
The implications of using alternate initial assumptions are explored in 
Section~\ref{paramimg}.
 
When comparing maps made in different colors, it is important to 
be mindful of changes in the performance of the imaging system.
Moving to longer wavelengths results in the loss of system resolution.
This is clearly demonstrated in the 3\,$\micron$ (PAHCS) image, which is 
consistent with having the same basic morphology seen in the 2\,$\micron$ 
images, however the lower resolution is not capable of revealing the fine 
detail.

The images at shorter wavelengths, in the J and H filters should, on the 
other hand, have higher resolution. 
Unfortunately, this theoretical improvement is not realized in the
maps we see. 
The main reasons for this are (1) the negative impact of the seeing 
becomes greater for shorter wavelengths; (2) the pixel scale of the
NIRC camera is not adequate to properly sample the finest interference 
fringes in this wavelength region; and (3) the target is relatively
dim, forcing the use of wide bandwidth filters making calibration more 
difficult \citep{keckmask}.
In the case of the J filter, no mask was used at all resulting in further 
noise from atmospheric turbulence.
These factors combine to result in a higher level of noise in the
reconstructed images, and a consequent loss of detail as compared with
the K-band maps.

The set of 2\,$\micron$ images (K, KCONT and CH4) in 
Figure~\ref{fig:sixcolour} present the opportunity to examine whether the 
bandwidth of the observation is a factor in determining the appearance of 
the image. 
These filters have similar center wavelengths (see Table~\ref{tab:filters}) 
however the fractional bandpasses are approximately 20\,\%, 7\,\% and 2\,\%  
for K, KCONT and CH4 respectively.
Some apparent differences were found to be due to changes with observing 
epoch (see Section~\ref{timev} below) as the K filter image of 
Figure~\ref{fig:sixcolour} differs by a couple of years from the others.
A more detailed quantitative examination, comparing only images from
the same epochs, revealed no systematic differences with filter bandwidth.
Indeed, this should not be surprising as there are no strong spectral
features that would imply a different origin for light isolated by these
three filters.

\subsection{Angular Sizes and Simple Brightness Models}
\label{simplemodel}

Although the images given in Figure~\ref{fig:sixcolour} do give a good
representation of the structure of \lkha, it is desirable to take a
step backwards towards the raw observables in extracting quantitative
information.
The prime reason for this is that fitting model brightness distributions
to the observed visibilities and closure phases is a far more direct
process than measuring intensities and sizes from an image reconstructed
by a complex nonlinear algorithm.

Furthermore, due to limited Fourier coverage, the mid-infrared 
interferometry from the ISI could not be inverted to produce a map
for comparison with those in Figure~\ref{fig:sixcolour}.
Information on the substantially cooler material to which
the ISI is sensitive needs to be incorporated through a process
of the fitting of simple model brightness profiles to the data.

The first set of fits made were to extract the position and relative flux 
of the $\sim$180\,mas binary companion in the near-IR data. 
The fitting algorithm employed both gradient descent and grid searching
to find the optimum chi-squared binary parameters from the 
calibrated visibility and closure phase data.
Relative fluxes  of the companion are given in the second column of 
Table~\ref{tab:models} for wavelengths in the K-band and shorter (the 
companion was not detected at long wavelengths) while separations are 
discussed later in Section~\ref{timev}.
The companion can be seen to be very much bluer than the resolved disk
component with its relative flux rising rapidly from K through to J-band.

Having established the parameters of the binary companion, it was 
possible to artificially remove it from the Keck data.
This was done using the simplistic approximation of a binary with two 
point components, however this proved easily adequate for the purpose of 
isolating the visibility function of the bright resolved disk.
Two-dimensional visibility data were then averaged azimuthally to enhance 
the signal-to-noise.

\begin{figure*}
\begin{center}
\includegraphics[angle=0,width=12cm]{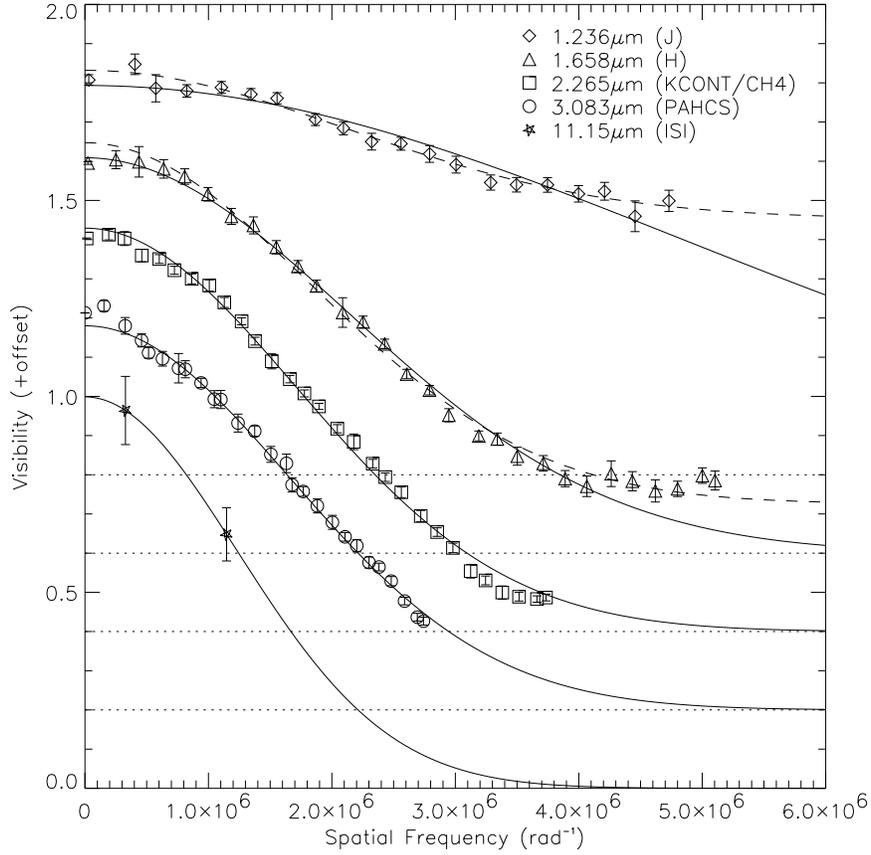} 
\caption{Visibility curves of \lkha\ from Keck aperture masking 
and ISI interferometry data.
Keck data has had successive visibility offsets of 0.2 added (horizontal
dotted lines) in order to separate the curves on the page.
The 1.236, 1.658 and 2.265\,$\micron$ curves have had excursions 
due to the presence of the binary companion removed (see text).
To enhance the signal-to-noise for the Keck data, curves depicted are 
the average of a number (typically 3--5) of separate observations taken 
at various epochs.
Overplotted solid lines show a best-fitting circular Gaussian function
to the data at each wavelength.
For the J and H filter observations, a more complicated 2-component model
is also shown (dashed line) which is discussed in the text.  }
\label{fig:viscurves}
\end{center}
\end{figure*}

Data at four wavelengths through the near-IR (Keck data) and at 
11.15\,$\micron$ (ISI data) are shown in Figure~\ref{fig:viscurves}.
Measurements from the entire course of the observing campaign have been
used, with a higher weighting given to the best epochs.
Visibilities were fit using simple circular Gaussian model
intensity profiles, with the full-width at half-maximum (FWHM) of
best fit given in the third column of Table~\ref{tab:models}.
Although the images of Figure~\ref{fig:sixcolour} show that the intensity
profile of \lkha\ is better represented by a disk with a central depression
than a Gaussian, we prefer here a simple one-parameter model to give
a robust estimator of the overall size. 
Despite its simplicity, the fits of this simple model in 
Figure~\ref{fig:viscurves} are quite good.

\begin{deluxetable}{cccccc}
\tablewidth{0pt}
\tablecaption{ {\rm Simple model fits to visibility data for each of
five different observing wavelengths (first column). 
The second column gives the fractional flux of the binary companion, while
the third column gives the best fitting FWHM of a Gaussian disk model of the
central resolved component.
The last 3 columns give fits of a more complex 2-component model to the
J and H-filter data only, as described in the text.
All errors were computed by observing the statistical spread in values found
by fitting the relevant models to separate, independently recorded data sets 
spanning a number of epochs.}
 }
\tablehead{
 \colhead{Wavelength} & \colhead{Binary Companion} & 
\colhead{Gaussian Disk} & & \colhead{Disk + Point Src} & \\
 & \colhead{Relative Flux} & \colhead{FWHM} & \colhead{Disk Flux} & \colhead{Disk FWHM}
 & \colhead{Point Flux}  \\
($\mu$m) & & (mas) & & (mas) & \\
}
\startdata
1.236 & 0.140$\pm$.021 & 16.1$\pm$0.8 & 0.31$\pm$0.03 & 36.8$\pm$3.7 & 0.55$\pm$0.03 \\
1.658 & 0.040$\pm$.013 & 36.4$\pm$0.8 & 0.84$\pm$0.02 & 42.1$\pm$2.4 & 0.12$\pm$0.02 \\
2.265 & 0.008$\pm$.003 & 43.7$\pm$1.8 &   &   &   \\
3.083 &      --        & 45.7$\pm$1.0 &   &   &   \\
11.15 &      --        & 62.6$\pm$10.9 &   &   &   \\
\enddata
\label{tab:models}
\end{deluxetable}

The most important thing to notice from the Gaussian disk fits of 
Figure~\ref{fig:viscurves} is the dramatically smaller size of the central 
disk in J-band. 
According to the FWHM sizes given in Table~\ref{tab:models}, \lkha\ shows a 
steady slow decrease in size over the large wavelength interval from 11.15 to 
1.658\,$\micron$, then a sudden halving between 1.658 and 1.236\,$\micron$.
Indeed, this dramatic reduction is also evident in Figure~\ref{fig:sixcolour} 
which shows the J-filter image having a barely-resolved core compared with 
images at other wavelengths.

Some increase in size with wavelength is to be expected from thermal considerations:
cooler dust further out will give a greater fractional contribution at long 
wavelengths. 
This probably explains the ($\sim$20\%) increase from 1.658 to 3.083\,$\micron$ 
and the ($\sim$27\%) increase from 3.083 to the ISI data point at 11.15\,$\micron$.
Although thermal radiative transfer modelling is required to fully exploit
the implications of these data for the structure of \lkha, the modest nature
of these increases seems to point to a fairly compact bounded structure whose
size is therefore not a strong function of observing wavelength.

The most logical explanation for the sudden drop in apparent size from 
1.658 and 1.236\,$\micron$ is that the central star itself is beginning
to outshine the disk at shorter wavelengths.
Visibility and closure phase data severely limit the range of allowed
locations of this stellar component to near the center of the resolved disk.
Placed anywhere else it would generate strong excursions which were not 
observed.

Further strong evidence for the presence of a hot (blue) point source comes from
examination of the H-filter visibility curve of Figure~\ref{fig:viscurves}.
At high spatial frequencies (above $\sim 4\times10^6 {\rm rad}^{-1}$)
the measured data departs systematically from the Gaussian disk fit in
a manner consistent with the presence of a weak point source.
This is illustrated with the dashed curve showing the unresolved component
holding up the visibilities at high spatial frequency.

Table~\ref{tab:models} shows results from fitting two component 
``Disk~+~Point Source'' models consisting of a point-component embedded in a 
Gaussian disk.
(Technically, these might be regarded as three-component models: recall that
the binary companion has already been subtracted from these data.)
This model was found to reduce the chi-squared misfit to the H-band visibility
data by an average factor of 1.9 over the seven epochs recorded, although this
improvement comes at the expense of an extra free model parameter.

The preferred model for \lkha\ can therefore be summarized as a resolved disk 
(approximated as a Gaussian of around 40\,mas FWHM) showing only a modest 
increase in apparent size across the entire near-IR, together with a distant
blue companion (contributing 14\% at J) and a bright embedded blue point 
source (contributing 55\% at J). 
Before proceeding with this interpretation, however, it is important to note 
that alternate scenarios might be indicated, particularly with the benefit of
full radiative transfer modelling and including more complex effects such
as scattering which may also play a role in altering the appearance of the
disk towards the blue.

In a recent imaging study of young massive stars, \citet{Leinert_01} reported 
the presence of an extended (0.9") halo contributing 0.15 of the flux in H-band. 
Evidence for this was not seen in the visibility curves of Figure~\ref{fig:viscurves}, 
where a sharp drop would be expected over the nearest few points from the origin.
Interferometric techniques (including both the results here and those of 
Leinert et al. 2001) can prove unreliable for the very shortest baselines 
(the so-called ``seeing spike problem''; e.g. Tuthill et al. 2000).
However, despite this, a halo as strong as 15\% would give signals which lie
well above this noise process and should have been seen here.
We hope that discrepancies such as this are sorted out as the techniques mature.

\subsection{Time Evolution of Images}
\label{timev}

\begin{figure}
\begin{center}
\includegraphics[angle=0,width=8cm]{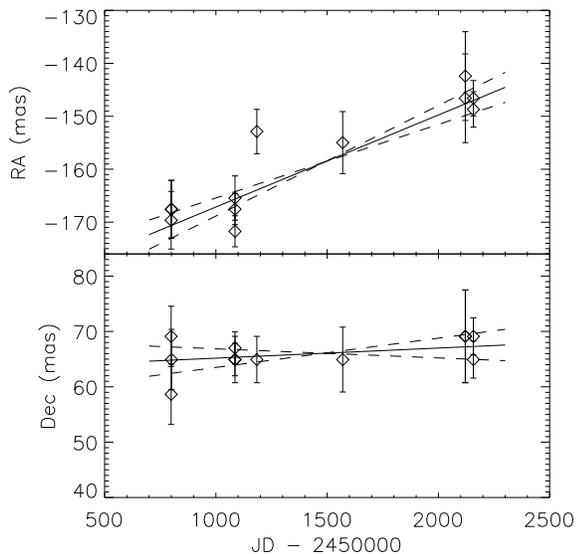} 
\caption{Relative position of LkH$\alpha$~101's companion as a function of
Julian date.  Upper panel gives Right Ascension, lower panel Declination.
The best fit uniform velocity model (solid line) is overplotted with the
one-sigma error in the component velocities indicated (dashed lines).}
\label{fig:orbit}
\end{center}
\end{figure}

With observations spanning almost four years, careful comparisons were made
in order to determine if \lkha\ exhibited significant changes of morphology
with time.
In particular, the location of the binary companion was carefully monitored
for evidence of orbital motion.
A plot of the separation (in right ascension and declination) is given in 
Figure~\ref{fig:orbit}.
The error bars are larger than can be obtained for simpler binaries consisting
of 2 point-source components due to ambiguity introduced by the need to 
locate a fiducial fixed point within the complex disk structure. 
For this purpose, the center of the circular disk (ignoring the asymmetric
limb-brightening) was chosen.

Figure~\ref{fig:orbit} shows the companion to be moving towards the West,
with almost all the change in the RA coordinate.
Fitting a uniform velocity to the data yields a motion of 6.4$\pm$1.3\,mas/yr 
at a position angle of 275$\pm$12\,degrees with a starting location of
182.5$\pm$2.4\,mas separation and position angle 69$\pm$2\,degrees in 1997~December~17
(JD 2450799).
Data are neither extensive nor precise enough to justify fitting of a full orbit 
at this stage.
The motion can be directly visualized from examination of the images presented
in Figure~\ref{fig:sixcolour}.
The J and K panels, whose data were taken in 2001~Sep, show the companion visibly
further to the West than the H, KCONT and CH4 images recorded in and around 1998.

Future monitoring of this companion to determine the orbit should help in 
pinning down the basic properties of the system, as discussed further in  
Section~\ref{basic_properties}. 
It will also be interesting to determine if the companion orbit is coplanar with 
the disk, as this has implications for the mechanism of formation of the binary 
system.
The interested reader is referred to the discussion in \citet{koresko98}, who 
found HK~Tauri~B to have a disk which was not coplanar with the binary companion.

\begin{figure*}
\begin{center}
\includegraphics[angle=0,width=14cm]{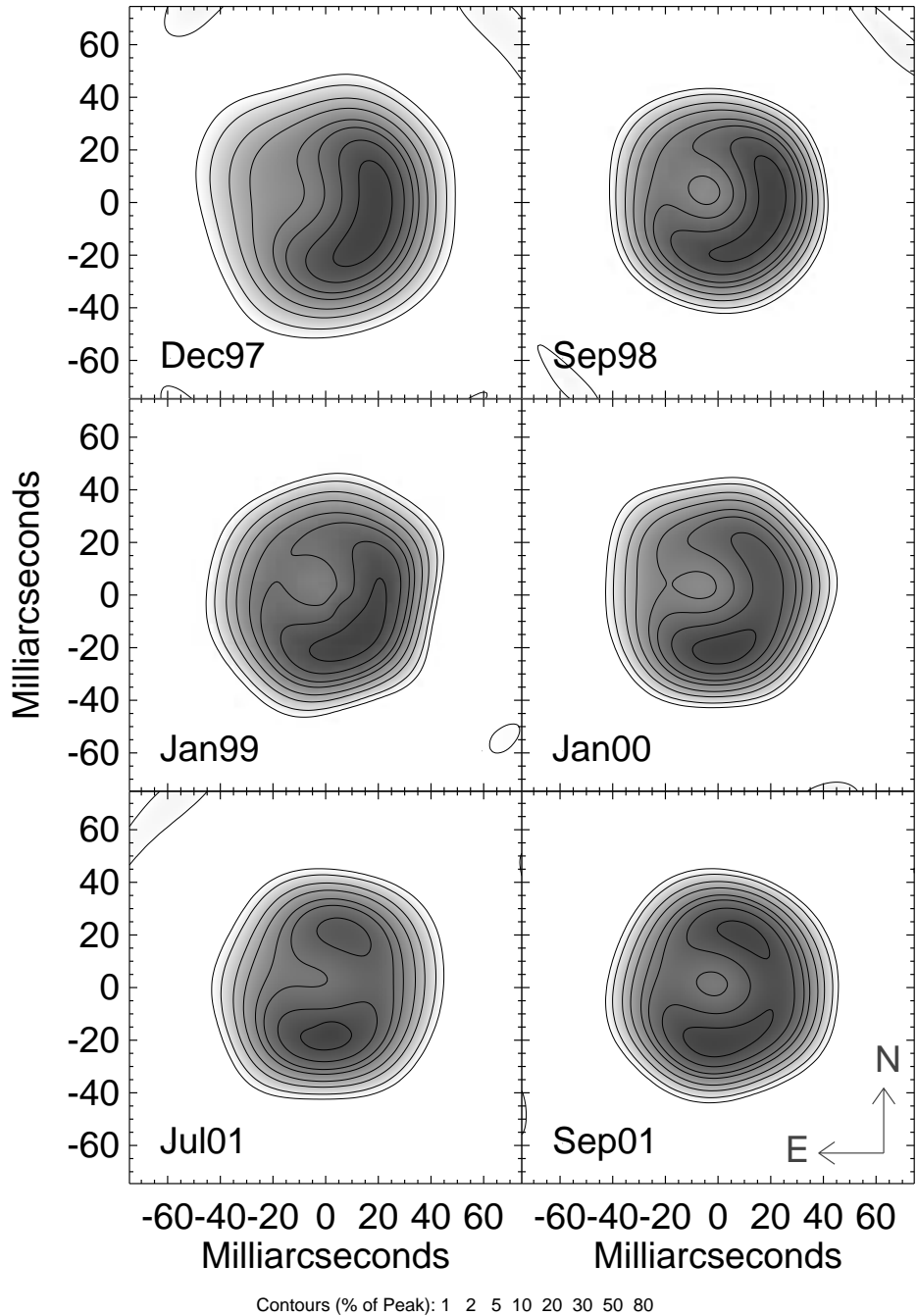} 
\caption{K-band images of the resolved disk of \lkha\ from data taken at six 
different epochs. With the exception of the KCONT filter which
was used in Dec97, all observations were made with the CH4 filter.}
\label{fig:evolution}
\end{center}
\end{figure*}

Systematic changes of structure were also recorded for the resolved disk
component of \lkha. 
This was studied by comparison of the highest fidelity images, which were all
produced in the K-band spectral region.
Images were superior as this represented the best compromise between
higher theoretical resolution (shorter wavelengths), more benign atmospheric
noise properties (longer wavelengths), and low contamination by the central
point-source (longer wavelengths).
Figure~\ref{fig:evolution} gives a series of K-band images from 1997~December 
through 2001~September depicting just the central disk component.

Although the basic structure of an asymmetrically limb-brightened circular disk 
with a central hole or depression is preserved in this series, there are more 
subtle changes associated with the peak intensity.
It can be seen that the brightest quadrant, initially to the west-southwest 
(1997/1998), gradually lengthens and shifts to the south (1999/2000), before
finally splitting into two: a northern and slightly brighter southern components (2001).
Although differences as large as this might be expected from noisy data in
any single map, this is unlikely to be the case here.
These changes are highly systematic over time, and multiple contemporaneous
observations at each individual epoch all corroborate the average map at that
epoch to a high degree. 

However, an important caveat on the interpretation of reconstructed structure
near the diffraction limit needs to be re-emphasized at this point.
Although we believe there has been real evolution of the morphology of 
LkH$\alpha$~101's resolved disk, the precise details of this change are unclear.
The Maximum-Entropy picture of the north-south migration and splitting of what 
was formerly a bright crescent to the west is only one of many possible valid 
models.
This question of ambiguity over the details of reconstructed images will be
discussed at greater length in Section~\ref{paramimg}.
However despite the uncertainties, the changes in the raw visibilities and 
phases which cause the evolution evident in Figure~\ref{fig:evolution} are real.

Photometry of \lkha\ could also be extracted from the data utilizing the
unresolved calibration stars as photometric references.
Although the experiment was clearly not optimized for such measurements,
we were able to establish that the infrared fluxes were constant to
within an uncertainty of 10\% over the course of the observing campaign.
Absolute flux levels were found to be in good agreement with published
photometry discussed in Section~\ref{sed}.

\section{Discussion}
\label{discussion}

\subsection{\lkha: a face-on accretion disk system}

The first qualitative impression given by the appearance of the images
of \lkha\ is that they conform well to our preconceptions for a young
system surrounded by an accretion disk.
The interpretation of the circular outline of the dust as a face-on view 
onto a disk, rather than a spherical structure, is well justified.
The inner regions have long been known to be highly inhomogeneous as 
testified by the bright reflection nebula NGC~1579 which must have a
low line-of-sight opacity to the central star. 
Further verification comes from studies of linewidths and excitation 
conditions \citep{Simon_Cassar_84} revealing the existence of a number
of regions.
A polar viewing angle for this star has also been favored in the literature
\citep{Barsony_90}.
The degree of circularity was measured by fitting to the images with the 
best signal-to-noise, and ignoring the brightest (asymmetric) structures.
No systematic elongations were found in any direction, enabling a limit
of $\simle 35 ^\circ$ to be placed on any tilt of the polar axis from the 
line of sight (assuming the disk itself to be circular and flat).

The marked central depression or gap evident in the images of 
Figures~\ref{fig:sixcolour}~and~\ref{fig:evolution} points to
a hole, or at least optically thin region at the center of the
disk surrounding the central star.
Such central cavities have long been invoked in order to help with
the fitting of the broadband spectra, and are also found in more recent
models \citep{Hollenbach_94,Yorke_96}.
The latter scenarios apply to massive stars upon attainment of their 
main-sequence luminosity, when they commence the erosion of their 
accretion disks in $\simle 10^6$\,yr with radiation and wind.
It is possible that our images of \lkha\ give the first direct view of 
this process at an early stage, with the star caught in this ephemeral 
phase having etched away the inner disk resulting in the dark hole at the 
center, but with substantial material still orbiting in a hot close ring. 

\begin{figure}[h]
\includegraphics[angle=0,width=8cm]{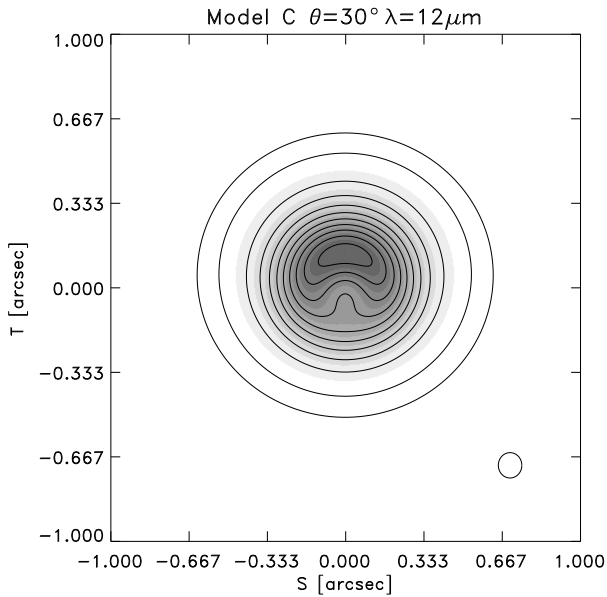} 
\caption{Model taken from \citet{Kessel_98} (Figure~7f) showing a 
simulated 12\,$\micron$ image of a photoevaporating disk, viewed at
an inclination of $30^\circ$.
\label{fig:kesselmodel}}
\end{figure} 

The final aspect of the image morphology which proves to be qualitatively
instructive is the asymmetric crescent-shaped brightening to the 
southwest, best illustrated in Figure~\ref{fig:evolution}.
The simplest interpretation is that the plane of the disk is tilted
{\em away} from our line of sight in the southwest, allowing a
clearer view to the hottest material on the inner wall facing
the star in this quadrant. 
Conversely, in the northeast side of the disk, being closer to us, 
cooler material further out is obscuring the hot inner parts.
Such geometrical arguments are only consistent with a thick disk,
torus, or strongly flared structure.
Thin flat disks, often encountered in the literature, will not produce
such effects.

The details of the appearance of a photoevaporating disk, including
a ``peculiar horseshoe-like feature'' for models with an inclination
of $30^\circ$, were predicted in simulations of \citet{Kessel_98}.
A model 12\,$\micron$ intensity map extracted from their paper is given 
in Figure~\ref{fig:kesselmodel} showing a strikingly similar morphology
to our images of \lkha.

The figure was included to illustrate the way in which a somewhat inclined,
centrally illuminated thick torus may make a good match to our observations.
Although such qualitative comparisons are highly encouraging, thick-disk
geometries with inner cavities are not unique to photoevaporating models
and establishing a physical model for this system will require more 
detailed quantitative analysis as discussed below.

\subsection{Basic properties}
\label{basic_properties}

Despite extensive observational study, there is a great deal of uncertainty 
over the basic properties of \lkha. 
A distance of 800\,pc, established by \citet{Herbig_71} from observations
of nearby stars thought to be in association, had become entrenched in
the literature.
However, this was recently challenged by \citet{Stine_Oneal_98} whose radio
photometry pointed to a much closer distance, suggesting that the star
may lie in an extension of the Taurus-Auriga star formation complex at
160\,pc.

\lkha 's bolometric luminosity, estimated at $4.8\ \times\ 10^4\ L_\odot$ 
\citep{Barsony_90}, was later revised downwards after it was found that
up to 75\% of the flux might be coming from a deeply-embedded cluster 
containing hundreds of objects \citep{Barsony_91}.
If the 160\,pc distance scale is adopted instead of 800\,pc assumed by
these earlier authors, the resultant luminosity of $4.8 \times 10^2 L_\odot$ 
would imply a mid--late B~star.
Although \lkha\ may not be as highly luminous as the O~star in earlier 
references, a mid--late B~star now seems to be an under-estimate.
Radio observations showing an ionized shell \citep{Harris_76,Cohen_82},
when corrected for a distance scale closer than 800\,pc,
find Lyman continuum photon fluxes as would result from
a significantly hotter early star \citep{Panagia_73}.
As the photoionizing flux is a strong function of spectral type, the radio 
photometry seems the most reliable indicator of the central star in this 
case.

Although revisions to the optical flux such as an intermediate distance 
scale and greater relative contribution from \lkha\ to the measured 
bolometric luminosity might bring it more into accord with the radio data, 
there are many difficulties such as known global anisotropies and uncertain 
optical depths complicating the picture.
However, we proceed with the assumption (based on matching radio photometry 
with expected Lyman continuum flux from given spectral types) that 
\lkha\ is an early B0$\sim$B0.5 zero age main sequence (ZAMS) star 
which implies an approximate mass $M_\star \simeq$ 10 - 20\,$M_\odot$; 
an effective temperature $T_{eff} \simeq 25\,000 - 30\,000$\,K; 
a luminosity $L \simeq 10\,000 - 25\,000$\,L$_\odot$; 
and a radius $R_\star \simeq 5$\,$R_\odot$   \citep{Panagia_73}.

The measured properties of the binary (apparent separation 180\,mas; 
motion 6.4$\pm$1.3 mas/yr) from Section~\ref{timev} allow us to set
fully independent constraints on the distance.
These are derived from the idea that the angular separation and velocity 
we see are related to the true parameters governing the binary motion 
(linear separation, semi-major axis, orbital velocity) by the distance 
combined with geometrical foreshortening factors.
Although we know neither these geometrical factors nor the true mass,
it can be seen below that estimates of distance can be discriminated
by making simple probabilistic assumptions about these unknown 
quantities (for example, that we are not looking from an unlikely, 
highly foreshortened perspective at the binary).

Firstly if we take the 800\,pc distance scale, then the orbital period 
of a binary with semi-major axis 180\,mas and an assumed system 
mass of 15\,$M_\odot$ is 446 years.
On the plane of the sky this would appear as a motion of 2.5\,mas/yr.
There are a number of ways our measurement of 6.4\,mas/yr 
could easily underestimate the true motion, however it is more
difficult to see how we have found a {\em larger} apparent motion.
The enclosed system masses would need to be 100\,$M_\odot$ in order 
to generate the observed apparent motion at this distance.
The companion could be approaching periastron in a highly eccentric
orbit and thereby moving faster than expected for its separation, or 
alternatively the objects may not form a bound pair at all in which 
case the apparent superposition we see will last $\simle 100$\,yr.
Although they cannot be ruled out, none of these explanations
seem very satisfactory and we conclude \lkha\ is likely closer
than 800\,pc.

If we consider now the 160\,pc distance, an apparent motion of 
28\,mas/yr (orbital period 40\,yr) would be seen for a 15\,$M_\odot$
circular binary in the plane of the sky with radius 180\,mas.
As mentioned above, there are a number of ways in which our 
measurements may underestimate the true state of affairs.
Apparent motions as low as 6.4\,mas/yr could be due to a large 
hidden line-of-sight component to the velocity vector, although
this is somewhat unlikely as it requires close alignment of the
present orbital direction to the line-of-sight.
The separation of 180\,mas used for the orbital computation might
similarly be an underestimate due to foreshortening (the companion
has a much larger separation, but on a vector close to our line-of-sight),
however this is statistically extremely unlikely, requiring much
closer alignments.
Allowing system masses as low as 5 and 10\,$M_\odot$ results in 
periods of 69 and 49\,yr giving apparent motions of 16 and 23\,mas/yr.
The 5\,$M_\odot$ assumption can match the observations without highly
unlikely inclinations, however it is doubtful that stars at this light 
end of this range are energetic enough to power the observed HII region.
As before, it is also possible to fit the observations by assuming 
a highly eccentric orbit: in this case the binary would need to have
been observed moving slowly near apastron. 

If we instead invert the question and ask what is the distance given 
the apparent motion, we arrive at 430\,pc for a 15\,$M_\odot$
circular binary orbiting in the plane of the sky.
However, we can do a little better than this, at least in a statistical
sense, by making the assumption that we have observed one-dimensional
projections of velocity and separation vectors taken from a random 
isotropic population (resulting in likely underestimation by a factor of 
$\pi/4$).
Accounting for this results in a likely distance of $\sim 340$\,pc, 
although values in the range 200--500\,pc can be comfortably accommodated
given the uncertainties in the mass and projection angle, and the 
experimental error on the apparent motion.
Allowing for eccentricity widens this range still further, although 
numerical simulations fitting for various orbits did find far more
solutions at the intermediate distance than at the extremes.
Unfortunately, the 180\,yr period of such a system makes it unlikely
that precise orbital elements, lifting some of the uncertainty in 
these calculations, can be derived soon.

In conclusion, although we can rule out neither of the earlier distance
estimates, our imaging results favor an intermediate $\sim 340$\,pc scale.
As the previous distance scales have not been established with any degree
of certainty, we include this new value despite the large uncertainties in 
its derivation above.
For the remainder of this paper, interpretations based on the short
\citep{Stine_Oneal_98}, intermediate (this work) and long \citep{Herbig_71}
scales are given.

\subsection{Spectral Energy Distribution}
\label{sed}

\begin{figure}[ht]
\includegraphics[angle=0,width=8cm]{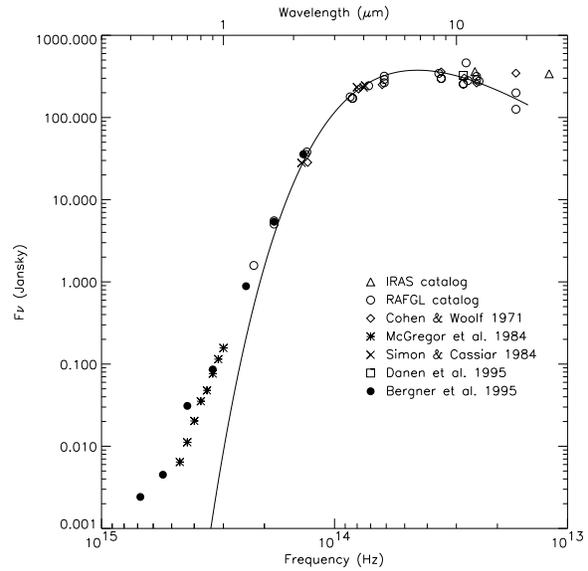}
\caption{Spectral energy distribution of \lkha\ from the optical through
the mid-infrared taken from various literature sources (as indicated). 
The overplotted solid line is a 750\,K blackbody curve 
\citep{Cohen_Woolf_71,DGB_95}
\label{fig:sed_raw}}
\end{figure}

The observed spectral energy distribution for \lkha\ is given in 
Figure~\ref{fig:sed_raw}, where the data have been taken from
various literature sources. 
The overplotted 750\,K blackbody curve was found 
\citep{Cohen_Woolf_71,DGB_95} to fit the near to mid-infrared spectral
regions.
We were able to extract photometry from our own observations, and found
the fluxes to be generally consistent with the literature values, 
in particular good agreement was found with the near-infrared data
of \citet{Bergner_95}.
We therefore presume that these historical data are fairly representative
of the present epochs under study.

An analysis of the energetics of \lkha\ and its associated region is 
complicated by the fact that it is known to lie behind thick clouds 
of obscuring material.
The spectrum must therefore be corrected for the wavelength-dependent
extinction (reddening).
This has been done using dust constants appropriate for so-called 
``outer cloud dust'' \citep{Mathis_90} expected in regions of heavy
obscuration.

Widely disparate estimates of the line-of-sight visual extinction ($A_v$) 
can be found in the literature, with values of 9.4, 9.7, 11.2, 14.2, 15.7, 
\& 18.5 obtained from, respectively, \citet{Barsony_90,McGregor_84,Kelly_94,
Thompson_77,Rudy_91,Hou_97}.
In the following discussion, we proceed with the relatively recent, 
intermediate value of 11.2 from \citet{Kelly_94}, although extinctions
encompassing the entire range were trialed.
When used to de-redden the spectrum of \lkha, values to the upper end 
of the range were found to give unrealistic spectra rising very sharply
in the blue.
At the other extreme (low $A_v$) the system could be modelled, but a
relatively small, cool central star was found making it difficult to 
reconcile with the radio measurements of Lyman continuum photon fluxes. 

\begin{figure*}
\begin{center}
\includegraphics[angle=0,width=14cm]{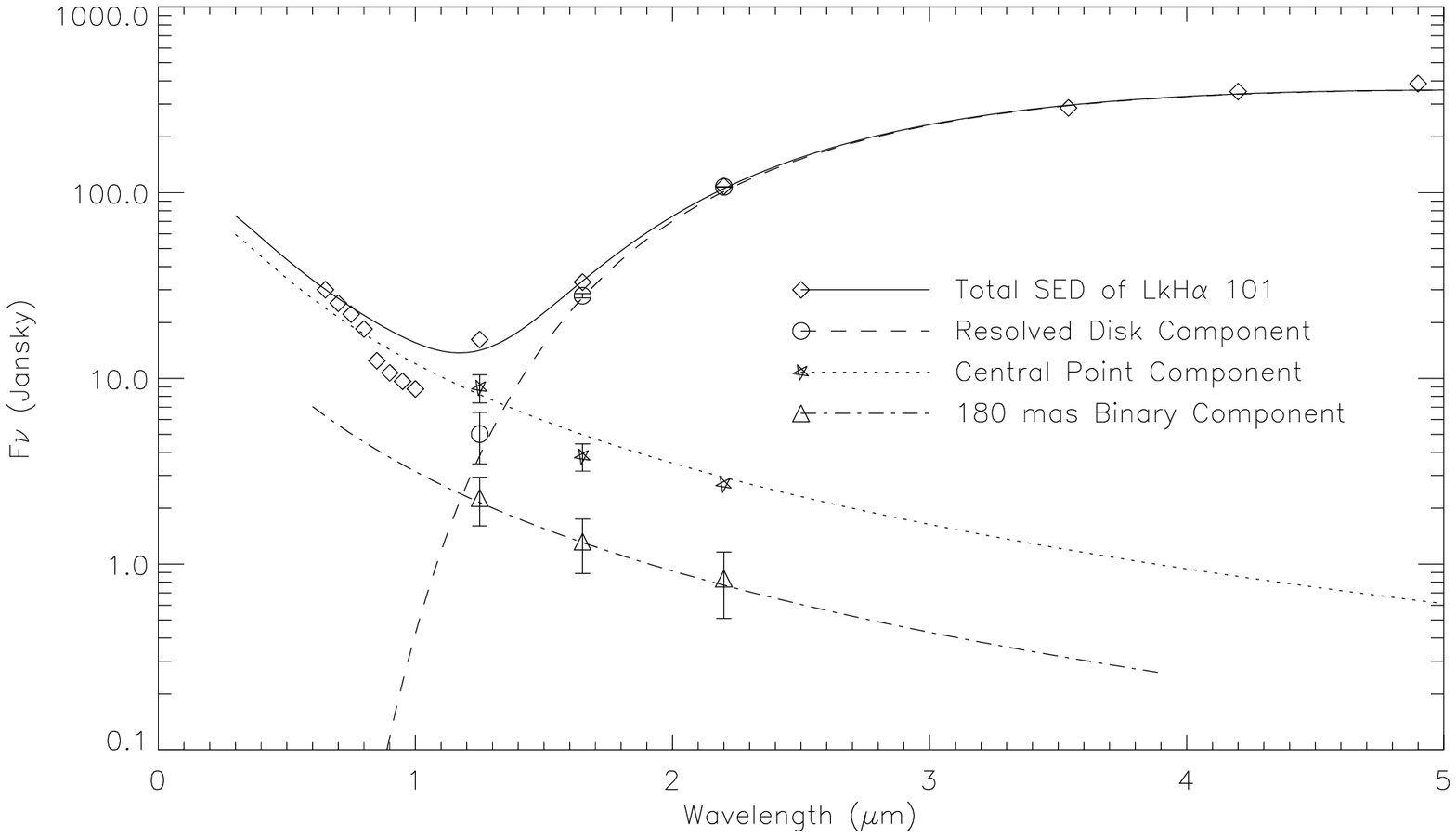}
\caption{Spectral energy distribution of \lkha\ corrected for 
wavelength-dependent extinction with $A_v=11.2$ 
\citep{Kelly_94}.
Using information from the imaging, the total SED (diamonds)
has been divided among the components of the \lkha\ system:
the binary companion (triangles), disk (open circles) and central
point source (stars). These separate components have also each been
modelled by blackbody spectra, as overplotted in the figure
(see text for details).
\label{fig:sed_fit}}
\end{center}
\end{figure*}

The de-reddened spectrum of \lkha, showing only the visible and 
near-infrared, is shown in Figure~\ref{fig:sed_fit}.
The spectral data points (diamonds) were taken from Figure~\ref{fig:sed_raw},
averaging multiple measurements at the same wavelength where necessary.
There is now a strong indication of a 2-component curve to the SED, 
with a red spectrum dominating wavelengths longer than K band, while
visible regions appear to betray the presence of a hot blue object.
The transition region between these two cases falls around 1\,$\micron$.

Utilizing information from the imaging and model fitting in Section~\ref{results},
we are able to decompose the full SED of Figure~\ref{fig:sed_fit} into separate
component spectra, each corresponding to individual features identified using
the high resolution techniques.
Firstly, we are able to extract the spectrum of the binary companion by
noting that we have obtained measurements of its relative flux in the
J, H and K bands as given in Table~\ref{tab:models}.
The resultant SED of the companion, isolated in this fashion, is plotted
as triangular points.

In a similar fashion, the remaining flux from the total can be divided into
components due to the disk and central point source, where the flux ratios
have again been taken from Table~\ref{tab:models}. 
The central point source, originally identified in Section~\ref{simplemodel}
from Figure~\ref{fig:viscurves}, is indicated by stars, while the resolved
component of the flux assumed to come from the disk is indicated by open circles.
Note that the central point component was only clearly identified at 2 
wavelengths, J and H, but we have also overplotted a central point 
component which contributes 2.5\% of the flux at K band. 
Although there may be weak evidence for such a signal in the visibility curve
of Figure~\ref{fig:viscurves}, this was done mainly as an illustration of
the expected K-band flux of the central component extrapolated from J and H.
For wavelengths longer than K, the circumstellar disk is expected to dominate
the SED.

For each of the three components of the system identified, a blackbody curve
has been plotted. 
Both stellar components have been modelled as having a temperature of 25\,000\,K,
(B0.5 ZAMS star from Panagia 1973) although the limited spectral coverage and 
uncertainties on the photometry mean that this is not well constrained by the data.
A near-infrared dust temperature of 1000\,K has been used to fit the 
resolved disk component, and in contrast to the stellar case, the temperature
is fairly well constrained to lie within $\pm 50$\,K of this value. 
The model spectral energy distributions for the central star, disk, and binary 
companion are given in Figure~\ref{fig:sed_fit} in dotted, dashed and dot-dashed
linetypes respectively.

Constraining the angular sizes to match the recorded flux levels, we deduce
that the dust shell must subtend an apparent area of 800\,mas$^2$ on the sky
(an area equivalent to a uniformly illuminated circle of radius 16\,mas).
This is in agreement, to within a factor of $\sim2$, with the flux-weighted 
observed angular size of the dust cloud ($\sim 1500$\,mas$^2$; radius of circle
of brightest emission 21\,mas) from the near-infrared imaging, and to within
a greater factor ($\sim4$) from the mid-infrared size.
These discrepancies could probably be taken into account by more complex 
models in which a variation in disk temperature and opacity with radius is 
allowed.

The central star and binary companion have apparent diameters of 56 and 29\,$\mu$as
(assuming the 25000\,K temperature).
Within the context of this model, we are able to analyze the energetics of the 
system, finding about $\sim 20$\% of the bolometric flux is reprocessed radiation 
from the dusty disk, while the binary companion contributes some $\sim 15$\%.
It is important to emphasize that these conclusions are dependent upon a
number of prior assumptions, for example the visible extinction and the 
temperatures of the stars (in particular, the temperature of the companion is 
all but unknown).
Although we have taken reasonable estimates for these quantities, the high
degree of uncertainty means that the resulting model should be regarded as a 
plausible exploration rather than an optimized global fit.

\begin{deluxetable}{lccc}
\tablewidth{0pt}
\tablecaption{ {\rm Derived physical parameters for the \lkha\ system, computed
for 3 different possible stellar distances.}  }
\tablehead{ }
\startdata
Distance & \colhead{160\,pc} & \colhead{340\,pc} & \colhead{800\,pc}  \\
$L_{star}$        & 1300\,$L_\odot$ & 5900\,$L_\odot$ & 30000\,$L_\odot$  \\
$R_{star}$        & 1.9\,$R_\odot$ &  4.1\,$R_\odot$  & 9.7\,$R_\odot$  \\
$L_{companion}$   & 340\,$L_\odot$ &  1500\,$L_\odot$ & 8500\,$L_\odot$  \\
$R_{companion}$   & 1.0\,$R_\odot$ &  2.1\,$R_\odot$  & 5.0\,$R_\odot$  \\
$L_{dust}$        & 270\,$L_\odot$ &  1200\,$L_\odot$ & 6700\,$L_\odot$  \\
$R_{dust}$        & 2.6\,AU & 5.4\,AU & 12.7\,AU  \\
$R_{cavity}$      & 3.4\,AU & 7.1\,AU & 16.8\,AU  \\
$R_{sublimation}$ & 2.5\,AU & 5.3\,AU & 12.5\,AU  \\
\enddata
\label{tab:properties}
\end{deluxetable}

It is now possible to use the prior estimates of stellar distance 
to translate from apparent to absolute quantities. 
This has been done in Table~\ref{tab:properties} for the three different
distance scales discussed in Section~\ref{basic_properties}.
The absolute luminosities (expressed in $10^3L_\odot$) and radii of the central 
source, the binary companion, and circumstellar disk are given on consecutive 
pairs of rows.
It is very interesting to note that our 340\,pc distance scale
results in a stellar luminosity and radius appropriate for a B1 ZAMS 
star \citep{Panagia_73}: in accord with our expectations from radio
photometry, and quite close to (although slightly smaller than) our initial 
model assumption (B0.5; $L=11000\,L_\odot$; $R=5.1\,R_\odot$). 
Note that neither the shorter nor the longer distance scales give this 
reassuring internal consistency for the model, however as noted above, 
different choices of $A_v$ or stellar temperature might plausibly result 
in models which work at these distances.

The value of $R_{dust}$ given is the diameter of a 1000\,K 
circular disk which would generate the observed blackbody flux
(note this is a simple circular area on the sky).
$R_{cavity}$ gives the expected linear size of the observed 21\,mas
radius circular ridge of brightest emission from the images of 
Figure~\ref{fig:evolution}. 
The close agreement between the observed and expected sizes (regardless of 
distance) gives confidence in the modeling of the SED and our choice of $A_v$.
It should be emphasized, however, that this is a very simple blackbody model
and a more detailed treatment should try to find a self-consistent thermal 
profile of the disk.

We identify the circular structure as the hot inner walls of the cavity 
whose size is likely set primarily by direct radiative dust sublimation from 
the central star (e.g. Tuthill et al. 2001) rather than disk-reprocessing or 
viscous-heating processes usually assumed 
\citep{Lynden-Bell_74,Hillenbrand_92}.
Finally, Table~\ref{tab:properties} gives $R_{sublimation}$, the 1500\,K 
sublimation radius of isolated dust grains for the system under the assumptions 
implicit for the 3 distance scales.
Grains surviving at the inner edge would be those best able to reradiate the 
stellar field, so the assumption of blackbody grains has been made.

It can be seen from the last two lines in the table that the sublimation 
radius is well matched to the size of the inner cavity, irrespective of the 
assumed distance.
This agreement is encouraging for a simple radiatively set inner cavity model, 
although clearly a more detailed treatment could also incorporate effects such 
as disk self-radiation \citep{Bell99}.
It is interesting to note that material must be continually evaporated at the 
inner edge in order to maintain this radius; this gas is presumably either
ultimately accreted by the central star or swept out into the surrounding 
HII region.

\subsection{Parametric Imaging}
\label{paramimg}

As we touched upon in earlier sections, the images produced by the Maximum
Entropy method are not unique, nor do the underlying algorithms always
represent the best possible approach. 
In particular, it may be possible to bring extra information to the 
mapping process, rather than making the assumption that the region of
sky to be imaged is most likely flat and featureless.

To take a concrete example, we can infer from Figure~\ref{fig:sed_fit}
that the images in the K-band should have a point source at the 
center contributing some $\sim2.5$\% of the flux.
Even concentrated into a single diffraction-limited beam, a 
feature as weak as this could not be seen amidst the bright structure 
within the resolved disk.
This is verified upon examination of images such as those in 
Figure~\ref{fig:evolution}, where no trace of the central point-source 
is seen.
However, even bright point sources can fail to appear in MEM maps as the
algorithm will penalize fine structure in the image if alternate 
smoother maps can be found which also fit the data.

\begin{figure}[ht]
\includegraphics[angle=90,width=8cm]{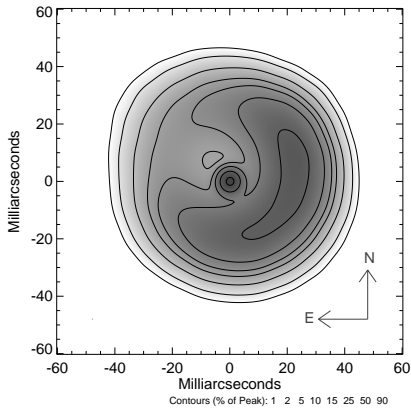} 
\caption{Image made with a 2.5\% point-source prior from data
taken in 1998~September through the CH4 filter.}
\label{fig:ptprior}
\end{figure}

In this case, we are able to bring our knowledge of the likely presence
of a point source into the mapping process.
This is done by using a {\em prior}, or default map (e.g. Monnier et al. 2001). 
The prior is the map the algorithm will default to in the absence of
any constraint from data. 
We have used a prior map which contains our expected 2.5\% point source
at the center, thereby permitting MEM to add fine structure without penalty.
The resultant image is given in Figure~\ref{fig:ptprior}.

The appearance of a compact central component, in accord with our
physical expectations for this system, teaches a valuable lesson in 
caution when interpreting maps made from data taken at close to the 
diffraction limit.
For partially-resolved targets, images cannot be definitive but have
some model-dependent structure.
However, this is not to say that the images are in any sense arbitrary.
A good example is the almost precisely central location of the point-source
in the Figure~\ref{fig:ptprior} image.
Although many trials with priors of various sorts were tried, all cases where
there was a strong point-source in the output map had it located at the center.
This not only agrees with our physical expectations for the system, it
is also required by the data. 
An off-center point source would generate strong telltale visibility 
and phase signals as was the case for the 180\,mas binary already mapped.

Features for which there is strong evidence in the data will always appear.
An example is the limb-brightened ring, which was present, almost
unchanged, for all images made.
More subtle things, such as the changes in the location of the brightest
part of the disk seen in Figure~\ref{fig:evolution}, are more open to
interpretation.
It is apparent that some form of systematic change in the brightness 
distribution is underway.
As the orbital period of material in the disk is about 5\,yr assuming the
340\,pc distance, it is tempting to interpret the changes as being
due to dynamical evolution of the disk.
Features such as spiral density perturbations were predicted by \citet{ARS89},
and such instabilities may be augmented by the presence of a binary companion, 
or accounting for self-gravity in a massive disk (e.g. Woodward et al. 1994).

However, the actual depiction in the figure of the bright western crescent
splitting to the north and south is model-dependent. 
For example, the location of the brightest quadrant on the disk was found 
to be affected by small variations in the location and brightness of the 
central star. 
This leads to the possibility that things unrelated to the disk itself, such
as varying line-of-sight extinction, may generate changes in appearance of
the reconstructed image. 
We refrain, therefore, from speculation on the detailed cause of these changes.

\section{Conclusions}
\label{conclusions}

Near-infrared images of the Herbig Ae/Be star \lkha\ have been obtained from a 
multi-epoch study utilizing interferometry on the Keck~1 telescope to obtain
information at the diffraction limit (tens of milli-arcseconds).
The mid-infrared U.C. Berkeley Infrared Spatial Interferometer was also able
to resolve the system at 11.15\,$\micron$. 
\lkha\ presents a resolved circular disk with a central hole or cavity.
A relatively blue-spectrum binary companion 180\,mas to the E-NE is also reported.

The morphology of the resolved limb-brightened ring is interpreted as a close 
to face-on viewing angle onto a geometrically thick torus or circumstellar
disk with a hot inner wall facing the central star.
The size of the cavity is consistent with the radiative equilibrium temperature
for dust sublimation.
A relatively slow increase in apparent diameter of the disk with observing 
wavelength over a decade from 1.2 to 11.15\,$\micron$ implies circumstellar 
material with a relatively compact density and/or thermal profile, arguing 
against classical power-law temperature profiles usually encountered in the 
literature.

Relative motion of the binary companion, together with simple assumptions 
on the masses and geometrical projection, favor a likely intermediate distance 
of around $\sim$340\,pc, although the earlier large (800\,pc) and particularly
small (160\,pc) suggested distance scales cannot be ruled out.

Combining the interferometry with published photometry enables a decomposition
of the spectral energy distribution into 3 component parts in the infrared. 
In addition to the 180\,mas companion and the resolved disk, we report the 
probable isolation of a bright unresolved source at the center of the cavity 
in the disk which has been identified as the primary.
This is likely an early B~star and the source of the photoionizing radiation
driving the HII region.
With more detailed radiative transfer modelling and studies at yet higher
resolution from the latest generation of optical interferometers, this
fascinating system is among the most promising candidates for in-depth study 
of the workings of a massive young star.

\acknowledgments

The authors would like to thank Rafael Millan-Gabet, David Hollenbach, 
Mary Barsony and Harold Yorke for helpful comments.
The image reconstructions presented here were produced by the
maximum-entropy mapping program ``VLBMEM,'' written by Devinder Sivia.
This research has made use of the SIMBAD database, operated at CDS,
Strasbourg, France, and NASA's Astrophysics Data System Abstract
Service.  Some of the data presented herein were obtained at the W.M. Keck
Observatory, which is operated as a scientific partnership among the
California Institute of Technology, the University of California and
the National Aeronautics and Space Administration.  The Observatory
was made possible by the generous financial support of the W.M. Keck
Foundation. The U.C. Berkeley ISI has been supported in part by grants from 
the Office of Naval Research (FDN00014-96-1-0737) and the National
Science Foundation (AST-9731625).

\end{document}